# Underwater Image Enhancement Based on Structure-Texture Reconstruction

Sen Lin, Kaichen Chi

*Abstract*— Aiming at the problems of color distortion, blur and excessive noise of underwater image, an underwater image enhancement algorithm based on structure-texture reconstruction is proposed. Firstly, the color equalization of the degraded image is realized by the automatic color enhancement algorithm; Secondly, the relative total variation is introduced to decompose the image into the structure layer and texture layer; Then, the best background light point is selected based on brightness, gradient discrimination, and hue judgment, the transmittance of the backscatter component is obtained by the red dark channel prior, which is substituted into the imaging model to remove the fogging phenomenon in the structure layer. Enhancement of effective details in the texture layer by multi-scale detail enhancement algorithm and binary mask; Finally, the structure layer and texture layer are reconstructed to get the final image. The experimental results show that the algorithm can effectively balance the hue, saturation, and clarity of underwater image, and has good performance in different underwater environments.

## I. INTRODUCTION

THE ocean is rich in resources, which is regarded as the "sixth continent" that can be used by human beings. As the main carrier of ocean information transmission, underwater image plays an important role in the exploration and development of the ocean. However, due to the complex underwater environment, the absorption and scattering of light by water bodies and suspended particles, there are problems such as color distortion and detail blur in underwater images [1], which makes it very difficult to extract image information accurately. Therefore, it is of great significance to get clear and real underwater images by appropriate methods.

To solve the above problems, many underwater image processing algorithms are proposed, including image enhancement methods based on the non-physical model and image restoration methods based on the physical model. In terms of enhancement: ANCUTI C O et al [2] used white balance and gamma correction respectively for the degraded image, and obtained clear image through multi-scale fusion; LI C et al [3] proposed a weakly supervised color transfer method to correct the color distortion of the underwater image; GUO Q et al [4] combined the dark channel segmentation pre-processing defogging algorithm and quantitative histogram stretching technology into the image enhancement process, effectively improving the clarity of the underwater scene; GAO Y et al [5] used the local triple fusion method based on the image formation model to fuse three images to get the final result. In terms of restoration: DAI C et al [6] selected background light through quadtree segmentation and scoring mechanism, decomposed the attenuation curve on RGB channel to calculate the transmission image, which can effectively restore degraded image; WANG B et al [7] simulated the visual processing of mammalian retina based on the retinal mechanism model, and realized the restoration of color distortion and contrast enhancement of underwater image; XIE H et al [8] estimated the optical parameters of water body needed for calculating the background light based on the optical theoretical formula, calculated the transmission function value by using the relationship between the scattering coefficient and the wavelength, and restored the underwater image by solving the imaging model inversely; DREWS et al [9] tested two hypotheses through experiments: the dark channel prior is still valid only when it is applied to the blue and green channels, and the blue and green channels of underwater images contain most of the visual information, to propose the Underwater Dark Channel Prior (UDCP) algorithm, which can effectively remove the uneven turbidity in the underwater image; WANG G et al [10] obtained the transmissivity of the backscatter component through the red dark channel prior and obtained the restored image by solving the dual transmissivity underwater imaging model inversely.

The above method is mainly to solve the problem of blur in the underwater image. Although the color correction algorithm is introduced, the effect in restoring the color balance is still not ideal, and there is too much noise in the underwater environment with high turbidity. In this paper, aiming at the color distortion caused by absorption and attenuation, further considering the effect of turbidity caused by scattering, an underwater image enhancement based on structure-texture reconstruction is proposed. The advantage of this algorithm is that the underwater image after color correction is decomposed into the structural layer and texture layer, and the transmittance map is robustly estimated from the structural layer without the interference of texture and noise. Through brightness and gradient screening and tone judgment, the background light estimation can be well controlled even under harsh imaging conditions to avoid the influence of white objects. In addition, the effective details in the texture layer are enhanced by the multi-scale detail lifting algorithm and the binary mask, and the structural layer and the texture layer are reorganized to obtain the final image. Compared with the existing algorithm, the results show that the algorithm can greatly improve the contrast and clarity of the image while effectively improving the color distortion of the underwater image, and the algorithm has good performance in different underwater environments.

## II. UNDERWATER IMAGING MODEL

The process of light propagation in water is similar to that in air. According to the Jaffe-McGlamery model [11], as shown in Fig.1,

Sen Lin is with the School of Automation and Electrical Engineering, Shenyang Ligong University, Shenyang 110016, China; State Key Laboratory of Robotics, Shenyang Institute of Automation, Chinese Academy of Sciences, Shenyang 110016, China; Institutes for Robotics and Intelligent Manufacturing, Chinese Academy of Sciences, Shenyang 110016, China. (e-mail: linsen@lntu.edu.cn).

Kaichen Chi is with the School of Electronic and Information Engineering, Liaoning Technical University, Huludao 125105, China.

the light received by the camera is usually composed of direct attenuation components, forward scattering component, and backward scattering component. The direct attenuation component is the part that the reflected light of the target object attenuates through the transmitting medium and received by the camera; The forward scattering component is the part of the reflected light of the object received by the camera after being scattered at a small angle; Different from the direct component and forward scattering component, the backscattering component does not come from the reflected light of the scene, but from the part of the camera where the background light is scattered by the suspended particles. The backscattering is the main reason for the decrease of the contrast and the color deviation of the underwater image. Generally, when the distance between the scene and the camera is small, the forward scattering component can be ignored. The simplified underwater imaging model is expressed as:

$$I_\lambda(x) = J_\lambda(x) \cdot t_\lambda(x) + B_\lambda \cdot (1 - t_\lambda(x)) \quad (1)$$

where $I$ is the underwater image, $J$ is the clear ideal image, $t$ is the transmissivity, $B$ is the global background light, $\lambda$ is any of the three RGB color channels of color image.

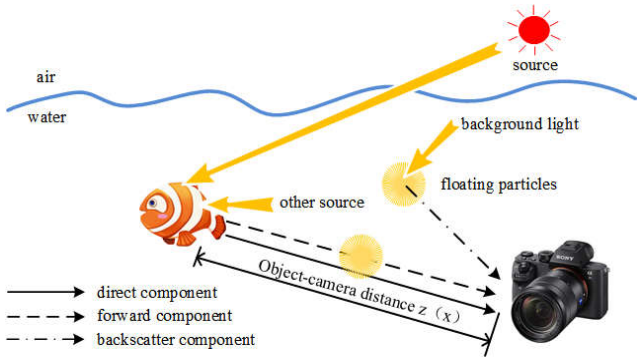

Fig. 1. Underwater optical imaging model.

## III. PROPOSED METHOD

### A. Color Correction

Because of the absorption and attenuation of light by water, the underwater images usually have serious color deviation. To solve this problem, an Automatic Color Enhancement (ACE) [12] algorithm is adopted, which mainly includes two parts: intensity adjustment and linear mapping.

1. The underwater color image is normalized so that the image can be zoomed in the domain $\Omega$ with a value of [0,1].
2. For the color normalized image, the RGB three color channels are adjusted by formula (2).

$$R(x) = \sum_{y \in \Omega \setminus x} \frac{S_\alpha(I_\lambda(x) - I_\lambda(y))}{\|x - y\|}, x \in \Omega \quad (2)$$

where $R(x)$ is the pixel value after intensity adjustment, $y \in \Omega \setminus x$ means $\{y \in \Omega : y \neq x\}$, $\|x-y\|$ is the euclidean distance between pixel $x$ and pixel $y$, $I(x)$ is the pixel value of degraded image $I$ at $x$, $I(y)$ is the pixel value of degraded image $I$ at $y$, $S_\alpha \in [-1,1]$ is a non-linear contrast adjustment transformation, it can enhance image details and suppress excessive edge information, the calculation method as formula (3).

$$S_\alpha(t) = \min\{\max\{\alpha t, -1\}, 1\}, \alpha \geq 1 \quad (3)$$

3. The value range of $R(x)$ is mapped to [0,1] linearly to obtain the enhanced color channel $R_r(x)$.

$$R_r(x) = \frac{R(x) - R_{\min}}{R_{\max} - R_{\min}} \quad (4)$$

where $R_{\max}$ is the maximum pixel value of $R(x)$, $R_{\min}$ is the minimum pixel value of $R(x)$.

Fig.2 is the contrast image and color histogram of the degraded image and the ACE algorithm.

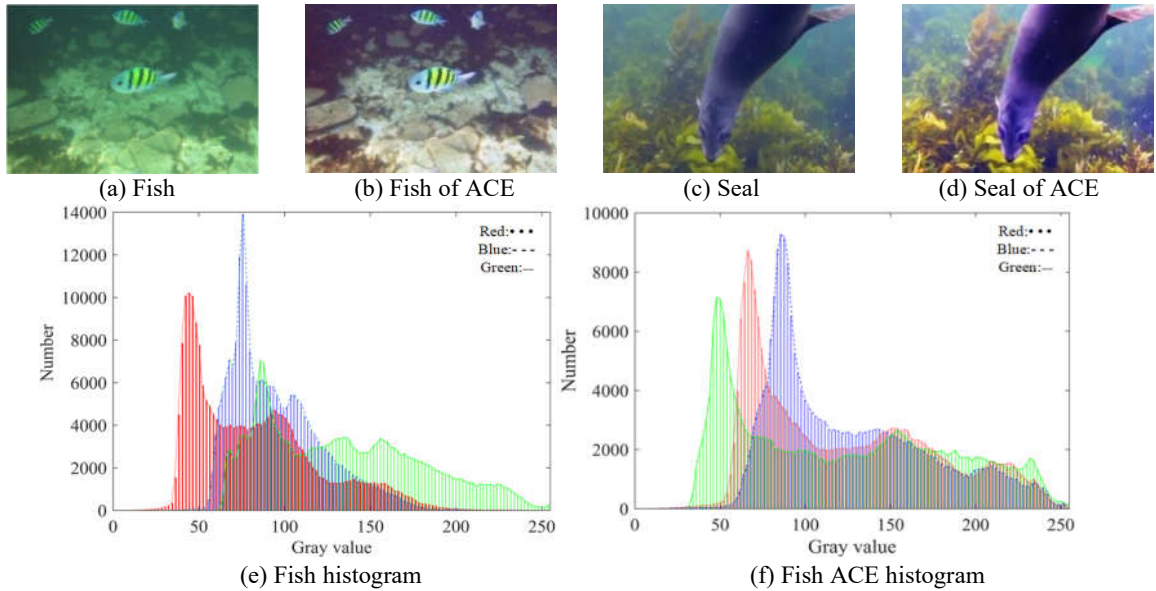

(a) Fish　　(b) Fish of ACE　　(c) Seal　　(d) Seal of ACE

(e) Fish histogram　　(f) Fish ACE histogram

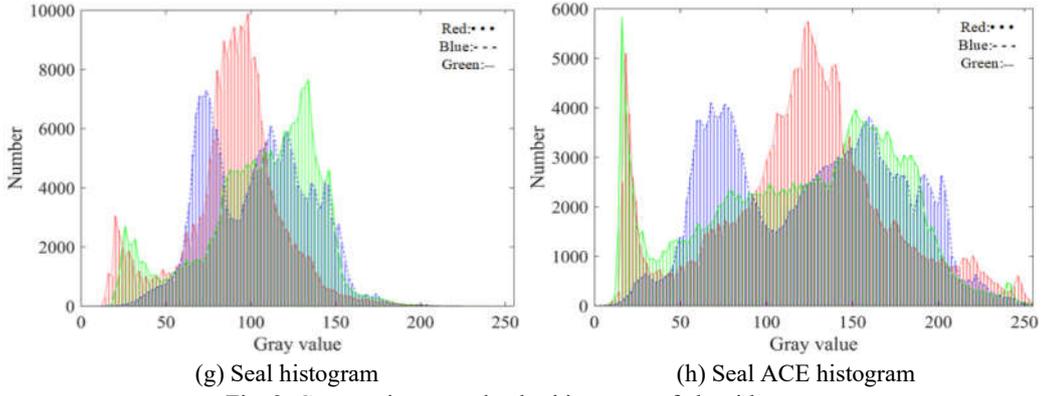

(g) Seal histogram          (h) Seal ACE histogram

Fig. 2. Contrast image and color histogram of algorithm.

It can be seen from Fig.2 that histogram (f) and histogram (h) compared with the other two histograms, the overall gray level is enhanced and the RGB color channel intensity distribution is more uniform. The image processed by the ACE algorithm has high color fidelity and good visual sense.

*B. Structure-Texture Decomposition*

To extract the structure of different types of degraded images, the Relative Total Variation (RTV) [13] model is introduced, which can effectively separate the structure and texture in the image. The model is as follows:

$$\arg\min_{S} \sum_{P}(S_P - R_r)^2 + \lambda \cdot \left(\frac{D_X(p)}{L_X(p)+\varepsilon} + \frac{D_Y(p)}{L_Y(p)+\varepsilon}\right) \quad (5)$$

where $R_r$ is the color correction image, $P$ is the index of 2D image pixels, $S$ is the output structure image, data item $(S_P - R_r)^2$ is helpful to reduce the separation error of structure image, $\lambda$ is the weight value used to control the smoothness of structure image, and the value range is [0.01,0.03] [13], $\varepsilon$ is used to prevent the denominator from being 0 and the value is $e-3$ [13], $D(p)$ is the total variation of pixel window, which can effectively distinguish complex texture types, $D_X(p)$ and $D_Y(p)$ are the total window variation of pixel $p$ in $x$ and $y$ directions, $L(p)$ is also the total variation of pixel window, which is slightly different from $D(p)$. This component is used to distinguish the prominent structure and texture information, $D_X(p)$, $D_Y(p)$, $L_X(p)$, $L_Y(p)$ expressions are as follows:

$$D_X(p) = \sum_{q \in R(p)} g_{p,q} \cdot |(\partial_X S)_q|$$

$$D_Y(p) = \sum_{q \in R(p)} g_{p,q} \cdot |(\partial_Y S)_q| \quad (6)$$

$$L_X(p) = \left|\sum_{q \in R(p)} g_{p,q} \cdot (\partial_X S)_q\right|$$

$$L_Y(p) = \left|\sum_{q \in R(p)} g_{p,q} \cdot (\partial_Y S)_q\right| \quad (7)$$

where $R(p)$ is a rectangular region centered on pixel $p$, $q$ is the index of all pixels in the range of $R(p)$, $\partial_X$ and $\partial_Y$ are partial derivatives in $X$, $Y$ directions, $g_{p,q}$ is a weighted function defined according to the degree of spatial approximation, expressed as:

$$g_{p,q} \propto \exp\left(-\frac{(x_p - x_q)^2 + (y_p - y_q)^2}{2\delta^2}\right) \quad (8)$$

where $\delta$ is used to control the size of calculation window, the empirical value is 5 [13], $x_p$ and $x_q$ are $x$ direction values of pixels $p$ and $q$, $y_p$ and $y_q$ are $y$ direction values of pixels $p$ and $q$.

The model described in formula (5) can still extract image structure effectively in a complex imaging environment and has strong robustness. The contrast image and transmission image of the degraded image and structure image are shown in Fig.3.

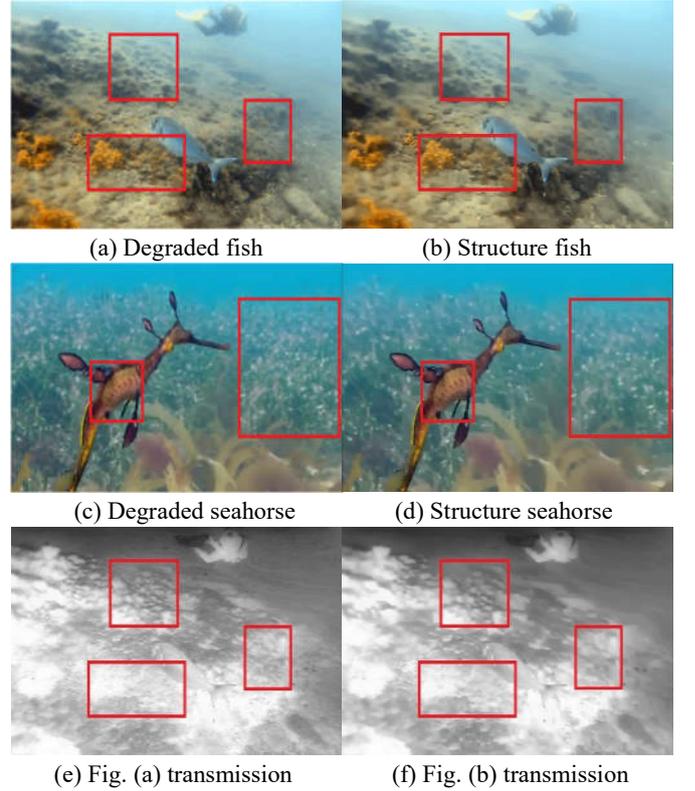

(a) Degraded fish     (b) Structure fish

(c) Degraded seahorse     (d) Structure seahorse

(e) Fig. (a) transmission     (f) Fig. (b) transmission

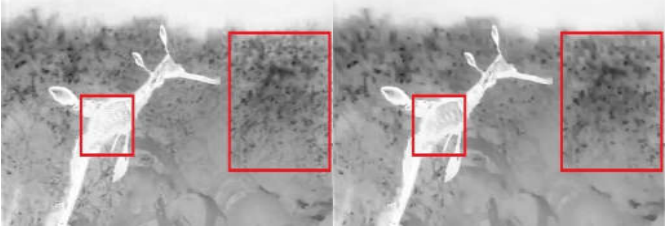

(g) Fig. (c) transmission      (h) Fig. (d) transmission

Fig. 3. Contrast image and transmittance image of degraded image and structure image.

It can be seen from Fig.3 that the texture information has been well separated from the structure contrast pictures (b), (d) and the transmission contrast pictures (f), (h), such as the water grass texture and the spots on the seahorse in the red rectangle. The model effectively retains the structure information of the degraded image.

*C. Structure Image Restoration*

The degraded image usually contains a lot of texture information and noise which can't be ignored. These interference factors will make the estimation of transmittance image unreliable, and then affect the restoration effect of image [14]. The denoised structure image obtained after the structure-texture separation has largely removed the influence of interference factors, and can well estimate the transmittance image.

Based on a large number of statistics on fog-free images, HE K believes that at least one of the three RGB color channels has a low-intensity value, even close to 0. This rule is known as the Dark Channel Prior (DCP) [15], which is mathematically defined as:

$$J_{dark}(x) = \min_{y \in \Omega(x)}(\min_{\lambda \in \{R,G,B\}} J_\lambda(y)) \to 0 \quad (9)$$

where $J_{dark}$ is a dark channel image, $J_\lambda$ are three color channels, $\Omega(x)$ is a local region centered on pixel $x$.

However, due to the attenuation effect of water on light, the attenuation degree of light with different wavelengths is different when it passes through the same distance underwater, resulting in a serious color deviation of underwater image. Therefore, the dark channel value estimated by formula (9) is small, and the transmissivity obtained later is large, so the transmission map solution in the prior algorithm of the dark channel is no longer applicable in the underwater environment. To meet the needs of underwater image restoration, a Red Dark Channel Prior (RDCP) [16] algorithm is introduced:

$$J_{RDCP}(x) = \min\left\{\min_{y \in \Omega(x)}[1-J_R(y)], \min_{y \in \Omega(x)}[J_G(y)], \min_{y \in \Omega(x)}[J_B(y)]\right\} \to 0 \quad (10)$$

where $J_{RDCP}(x)$ is the prior value of red dark channel, using $[1-J_R(y)]$ instead of $J_R(y)$ can effectively avoid the serious color deviation caused by red light attenuation.

Divide $B_\lambda$ into both sides of formula (1) and substitute formula (10) to find the transmission of the backscattered component:

$$t(x) = 1 - \min\left\{\frac{\min_{y \in \Omega(x)}[1-I_R(y)]}{1-B_R}, \frac{\min_{y \in \Omega(x)} I_G(y)}{B_G}, \frac{\min_{y \in \Omega(x)} I_B(y)}{B_B}\right\} \quad (11)$$

The selection of the underwater image background light will be interfered with by the white object so that the area with the highest brightness in the image is no longer the background light area. According to the characteristics of high brightness and flatness of background light, a robust background region is obtained. Firstly, the brightness channel and gradient images of the underwater image are binarized to get high brightness region and flat region, and the intersection region of them is obtained. The high-luminance area of the image is determined by the formula (12):

$$H_t(x) = \begin{cases} 0, H(x) < \alpha \\ 1, H(x) \geq \alpha \end{cases} \quad (12)$$

where $H$ is the luminance component of LAB color space, $H_t$ is the binary result of $H$ component, the threshold $\alpha$ value is:

$$\alpha = H_{mean} + (255 - H_{mean})/3 \quad (13)$$

where $H_{mean}$ is the mean value of $H$ component, use the gradient map $T(x)$ of the luminance component $H$ to determine the flat area of the image:

$$T_t(x) = \begin{cases} 1, T(x) < \beta \\ 0, T(x) \geq \beta \end{cases} \quad (14)$$

where $T_t$ is the binary result of gradient map, and the threshold value $\beta$ is:

$$\beta = \min(U_T, \gamma) \quad (15)$$

where $U_T$ is the gradient corresponding to the maximum probability in histogram distribution of gradient map $T$, $\gamma$ is the empirical value, set to 20 [17].

The intersection area of pixels satisfying both $H_t(x)=1$ and $T_t(x)=1$ is the preliminary estimated background light point, as shown in Fig.4 (b). The underwater light attenuates in a wavelength-dependent way, and the red light attenuates fast, so the underwater image usually has a blue-green tone, which means the blue-green component of the background light is higher. White interfering pixels have the characteristic that their intensity is high in all three color channels. To further avoid extracting white objects in the intersection area, the best background light point is selected by the judgment of hue. First, the mean value $I_{mean}^R$, $I_{mean}^G$, $I_{mean}^B$ of RGB channel is obtained, if $I_{mean}^B \geq I_{mean}^G$, the image is blue tone, and the pixel with the largest gray difference between blue and red channel is selected as the candidate background light point, if $I_{mean}^B < I_{mean}^G$, the image is green tone, and the pixel with the largest gray difference between green and red channel is selected as the candidate background light point. Finally, the first 0.1% of the pixels with the highest brightness are selected as the water illumination value $B_\lambda$. The algorithm steps are as follows: get the gray histogram $h$ of candidate background light points, accumulate the histogram $h$ from right to left to get $h_{sum}$, when $h_{sum} > 0.1\% \cdot N$, $N$ is the total number of pixels, and get the histogram abscissa

value z here. Obtain the background light value of the water body in the range of [z,255]:

$$B_\lambda = \frac{\sum_{i=z}^{255} i \cdot h(i)}{0.1\%N} \quad (16)$$

After calculating the transmittance and the background light value of the water body, the turbidity removal image can be recovered from formula (1):

$$J_S(x) = \frac{S(x) - B_\lambda}{\max[t(x), t_0]} + B_\lambda \quad (17)$$

where $t_0$ is a critical value set to avoid $t(x)$ too small, it can effectively prevent the restored image from appearing over bright pixel points or pixel areas, with a value of 0.1 [15]. The comparison of experimental results is shown in Fig.4.

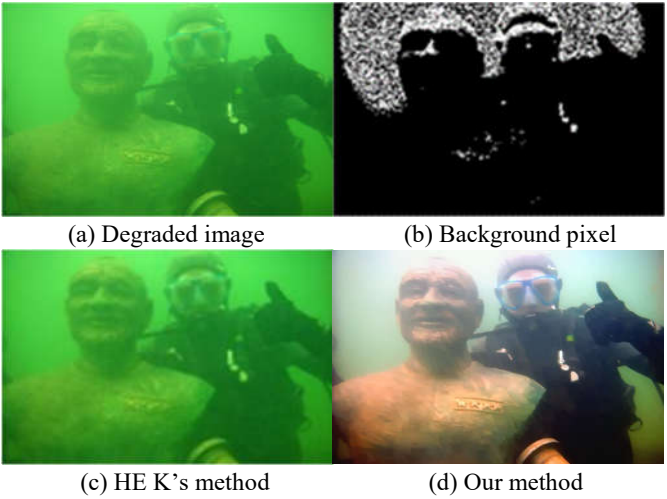

(a) Degraded image  (b) Background pixel
(c) HE K's method  (d) Our method
Fig. 4. Comparison of experimental results.

### D. Texture Image Enhancement

The texture image $W$ is obtained by subtracting the structure image $S$ from the color correction image $R_r$:

$$W = R_r - S \quad (18)$$

Firstly, the whole details of texture image $W$ are enhanced by multi-scale detail lifting algorithm, this algorithm applies Gaussian kernels of different scales to $W$ to obtain three different levels of blurred images:

$$B_1 = G_1 * W \quad B_2 = G_2 * W \quad B_3 = G_3 * W \quad (19)$$

where $G_1$, $G_2$, $G_3$ is a standard Gaussian kernel, the deviation $\sigma$ determines the smoothness, the larger $\sigma$, the better the smoothness of the Gaussian filter, therefore, in order to obtain three different levels of blurred images, the deviation value is $\sigma_1 = 1.0$, $\sigma_2 = 2.0$, $\sigma_3 = 4.0$.

Extract fine details $C_1$, middle details $C_2$ and rough details $C_3$ of image:

$$C_1 = W - B_1 \quad C_2 = B_1 - B_2 \quad C_3 = B_2 - B_3 \quad (20)$$

Merge three levels of detail to generate an overall detail image:

$$C = [1 - \omega_1 \cdot \text{sgn}(C_1)] \cdot C_1 + \omega_2 \cdot C_2 + \omega_3 \cdot C_3 \quad (21)$$

where $\omega_1$, $\omega_2$, $\omega_3$ are 0.5, 0.5, 0.25 [18] respectively. Excessive fine details $C_1$ increase the gray difference near the edges, which may saturate the edges. In order to solve this problem, the coefficient of $C_1$ is controlled by the symbol function in formula (21) to keep the average difference of fine details between different scales, it can effectively enhance image details while suppressing saturation.

Secondly, the enhanced texture image $C$ may still contain unwanted residual details in the smooth region, therefore, the binary mask $M$ is introduced to divide the texture image into smooth region and detail region, and effectively remove the interference artifacts in the smooth region. Discrete Cosine Transform (DCT) coefficients are used to check whether the image is smooth. Each $8 \times 8$ matrix in the image area defines its DCT coefficient as $E$. The possibility of residual details in the area is defined as:

$$p = \sum_{x,y} E_{x,y}^2 - E_{1,1}^2 - E_{1,2}^2 - E_{1,2}^2 \quad (22)$$

where $x, y$ is the coordinate, if $p > 0.1$, $M$ is set to 1, if $p \leq 0.1$, $M$ is set to 0.

As a result, a texture image $J_C$ with enhanced detail and elimination of artifacts can be obtained. The comparison of the texture image before and after enhancement is shown in Fig.5, in this article, we magnify the texture 10 times for better visualization.

$$J_C(x) = M \cdot C(x) \quad (23)$$

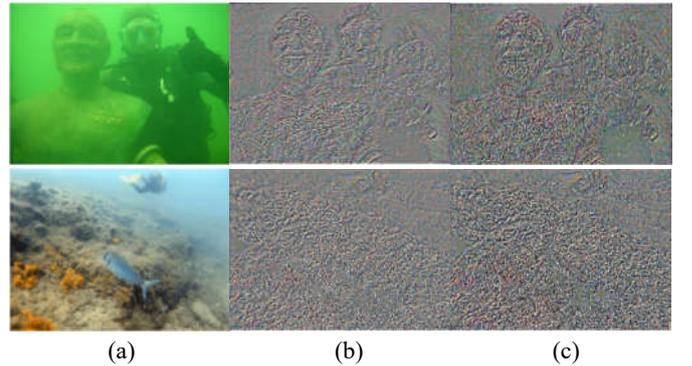

(a)   (b)   (c)
Fig. 5. Comparison of experimental results: (a) Degraded image (b) Texture images (c) Enhanced texture images

It can be seen from Fig.5 that picture (c) has effectively enhanced the image details compared with picture (b), for example, the details of the diver's body and the texture of the underwater fishes and plants in the picture (c) are clearer.

### E. Structure-Texture Reconstruction

Add the enhanced texture image and the restored structure image to get the final result image $J$.

$$J(x) = J_s(x) + \tau \cdot J_c(x) \qquad (24)$$

where $\tau$ is the scale factor and the value is $1/t(x)$ [14].

*F. Algorithm Flow*

In this paper, the ACE algorithm is first used to correct the color of the degraded image. Then, the relative total variation model is introduced to decompose the color balanced image into structural layers and texture layers. After that, an improved dark channel prior algorithm is proposed to remove the turbidity of the structural image. The detail enhancement algorithm and binary mask enhance the effective details of the texture image. Finally, the enhanced structure layer and texture layer are recombined to obtain the final image. The algorithm pseudo-code is shown below, Fig.6 is the algorithm flow of this paper.

---

**Algorithm:** Underwater image enhancement based on structure-texture reconstruction

**Input:** Degraded image

**Output:** Result image

1. Correct the degraded image color using Eq. 4;
2. Color correction image is divided into structure layer and texture layer using Eq. 5;
3. Restore structure image using improved dark channel prior algorithm;
       Calculate backscatter component transmittance using RDCP algorithm;
       Select best background light based on brightness, gradient discrimination and hue judgment;
       Plug underwater optical imaging model to obtain structure restoration image;
4. Enhance texture image;
       Enhance overall details of texture image using Eq. 21;
       Remove interference artifacts in smooth regions using Eq. 23;
5. Add the enhanced texture layer to the restored structure layer to get the final result image;

---

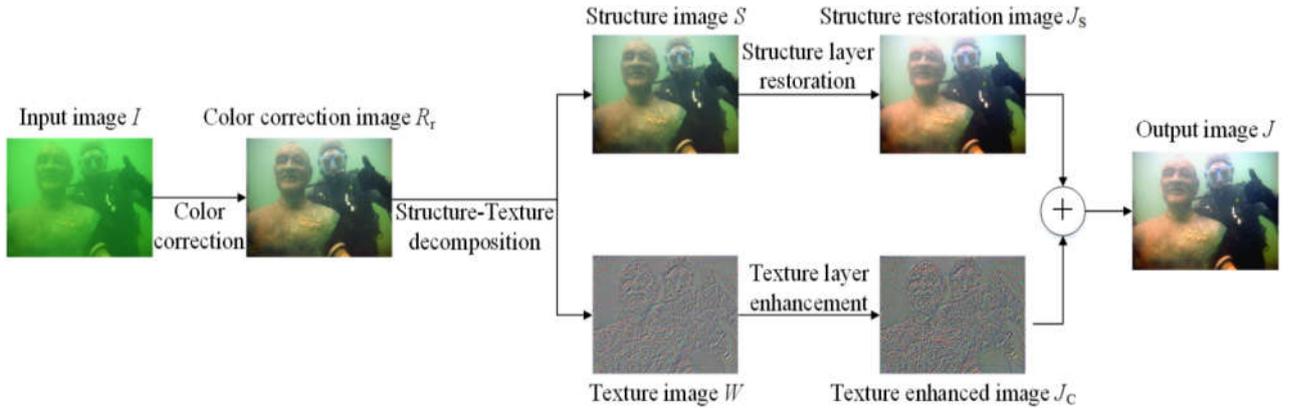

Fig. 6. Algorithm flow.

## IV. EXPERIMENTAL RESULTS AND ANALYSIS

To verify the effectiveness of the algorithm, the result images in different environments are compared and analyzed from subjective evaluation and objective evaluation. In the aspect of subjective evaluation, the representative images were selected from the underwater image depot [19], [20] established by JIAN M and MA Y in recent two years. In the aspect of objective evaluation, it is difficult to obtain the underwater reference image in an ideal state, so we use four non-reference evaluation indexes are selected for quality evaluation to ensure the authenticity and accuracy of the objective evaluation, and the algorithm in this paper is compared with other algorithms. Finally, through the experiment of feature point matching, it is further verified that our algorithm can play an important role in texture detail enhancement.

*A. Subjective Evaluation*

1. Color correction effect evaluation

To evaluate the effectiveness of this algorithm in color correction, a group of underwater natural scenes is selected for histogram distribution analysis, the underwater natural scene and histogram comparison image are shown in Fig.7.

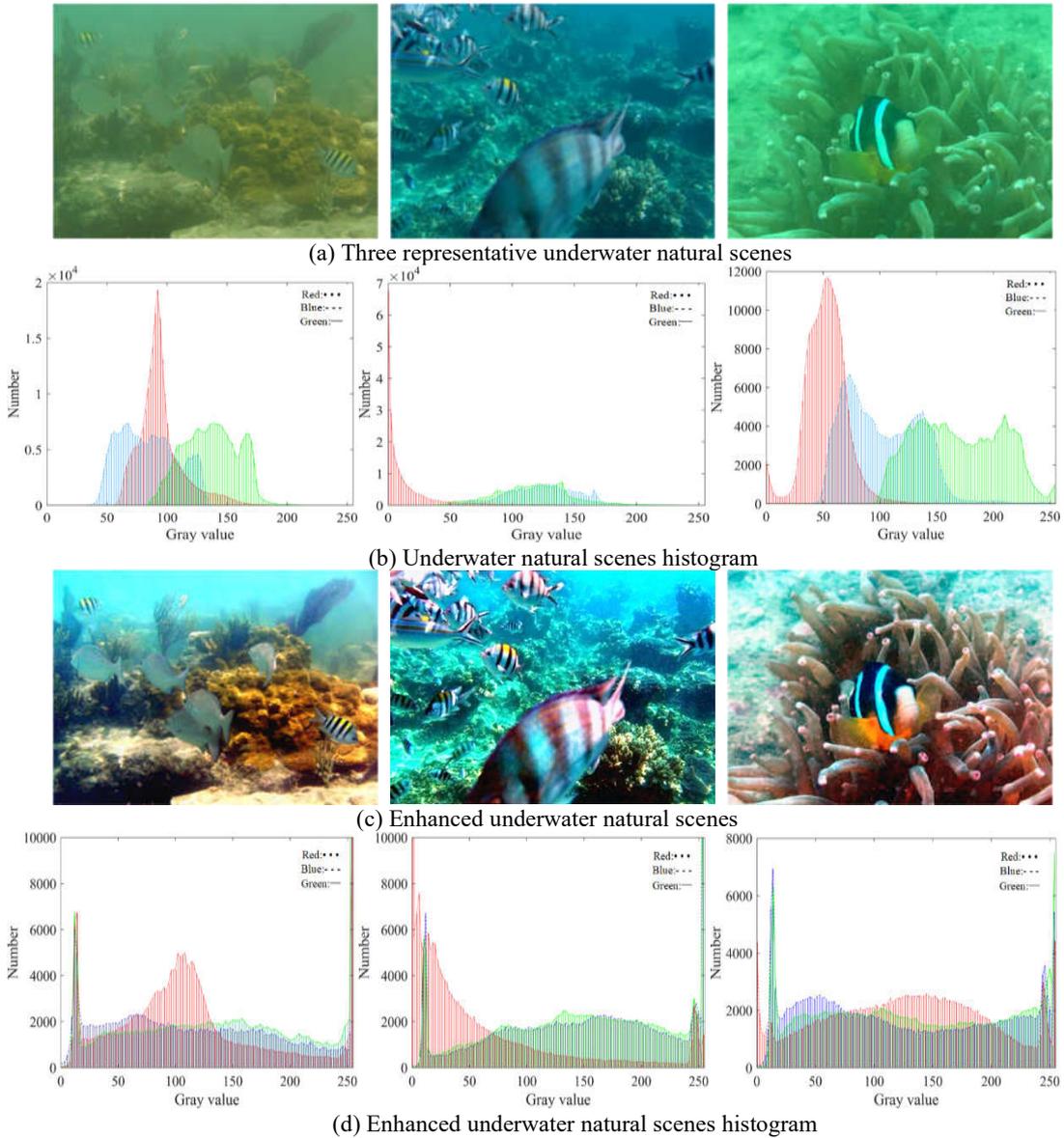

(a) Three representative underwater natural scenes

(b) Underwater natural scenes histogram

(c) Enhanced underwater natural scenes

(d) Enhanced underwater natural scenes histogram

Fig. 7. Contrast image and color histogram of underwater natural scenes.

It can be seen from the picture (d) that the intensity distribution of RGB three channels of the enhanced color histogram is more uniform. In picture (c), the color of the enhanced underwater natural scene is bright and the brightness distribution is even. For example, the color of coral and fish becomes richer and no longer presents the blue-green caused by the phenomenon of color deviation. The algorithm has achieved good results in color correction in the natural scene.

Next, the effectiveness of the algorithm in multi-color system color correction is tested through color card recovery experiments. This experiment is based on the undistorted color card image, which has color degradation due to the complexity of the underwater imaging environment. By processing the degraded image with the algorithm, the color recovery effect of the algorithm can be effectively verified. The algorithm in this paper is compared with the classic Underwater Dark Channel Prior (UDCP) [9] algorithm and the latest fusion [2] algorithm, Image Color Correction Based on Double Transmission Underwater Imaging Model [10] algorithm, Deep Underwater Image Enhancement Network (DUIENet) [21] algorithm, Underwater Image De-Scattering and Enhancing Using DehazeNet and HWD (DehazeNet and HWD) [22] algorithm, the color recovery experimental comparison image is shown in Fig.8. Besides, the CIEDE2000 evaluation index [23] quantifies the color difference between the standard color card and each processed color block, the value range of CIEDE2000 is [0,100]. The smaller the value, the smaller the color difference. Table I shows the CIEDE2000 evaluation value of the first experimental image in Fig.8, and the bold font is the optimal value of the corresponding algorithm.

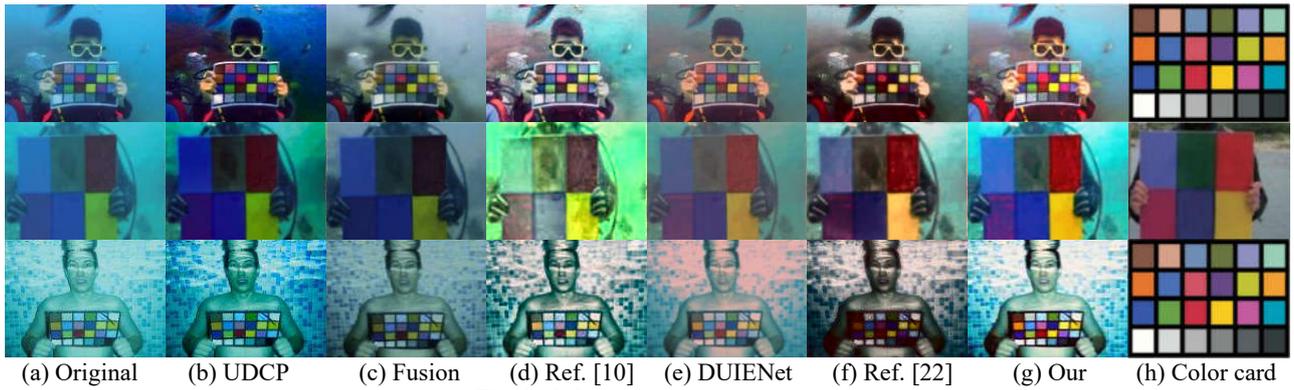

(a) Original  (b) UDCP  (c) Fusion  (d) Ref. [10]  (e) DUIENet  (f) Ref. [22]  (g) Our  (h) Color card
Fig. 8. Color recovery experiment.

It can be seen from Fig.8 that the color of multiple color blocks in the color card image processed by the UDCP algorithm deviates, the contrast of similar color blocks is low, such as yellow and light green, and the overall color of color card image is deepened. The yellow and light green color blocks processed by the fusion algorithm have low discrimination, and the light blue color block appears distortion, and the color of the color card is not bright. The Image Color Correction Based on Double Transmission Underwater Imaging Model [10] algorithm has an obvious color deviation in the second group of experiments, and the experimental results are not robust. The color image of the DUIENet algorithm is overall reddish. In the third group of experiments, obvious red artifacts appeared, and the color correction effect was significantly different from the real color card. Dehazenet and HWD [22] algorithm is more accurate for color correction, but the brightness of the color card image is reduced, and the visual effect of each color block is not consistent with the real color card. The color card image of the algorithm in this paper has vivid colors, the color of each color system can be distinguished, and the resulting image color is close to the real color card. The color contrast of similar color systems such as yellow and light green, blue and purple is enhanced.

TABLE I.  Evaluation of CIEDE2000 index

| Method | | | | | | | | | Avg |
|---|---|---|---|---|---|---|---|---|---|
| UDCP | 16.00 | 32.19 | 30.84 | 11.92 | 34.26 | 14.71 | 30.81 | 24.36 | 24.03 |
| | 15.59 | 18.91 | 11.53 | 23.39 | 18.31 | 10.11 | 27.91 | 19.40 | |
| | 29.52 | 39.53 | 6.43 | 30.84 | 42.74 | 36.13 | 29.12 | 22.13 | |
| Fusion | 20.73 | 23.28 | 16.18 | **5.86** | 25.58 | 7.94 | 32.32 | 10.20 | 16.90 |
| | 14.59 | 14.65 | **9.32** | 22.17 | 8.92 | 8.87 | 30.72 | 14.35 | |
| | 28.22 | 24.92 | 8.34 | 13.94 | 19.57 | 19.39 | 14.35 | **11.10** | |
| Ref. [10] | 17.47 | 17.03 | 13.47 | 30.63 | 5.65 | 11.97 | 19.75 | 8.62 | 14.61 |
| | 10.25 | 14.54 | 10.65 | 13.56 | 8.37 | 13.11 | 28.15 | 11.49 | |
| | **9.50** | **19.80** | 7.64 | 15.89 | 11.81 | **12.64** | 20.32 | 18.35 | |
| DUIENet | **5.32** | 13.80 | 12.15 | 8.62 | 12.86 | 11.11 | 11.11 | 12.00 | 12.67 |
| | 9.10 | **9.23** | 9.33 | **8.20** | 9.65 | 8.68 | 14.85 | 9.97 | |
| | 12.76 | 24.26 | 21.73 | 20.09 | 10.49 | 19.03 | 14.27 | 15.50 | |
| Ref. [22] | 14.87 | **10.93** | 18.90 | 8.61 | 18.42 | **5.16** | 12.80 | 25.07 | 17.70 |
| | 12.80 | 22.39 | 9.73 | 11.60 | 21.93 | **4.47** | 27.69 | **4.99** | |
| | 27.22 | 53.54 | 14.81 | **11.30** | 9.54 | 30.35 | 25.62 | 22.00 | |
| Our | 13.38 | 13.82 | **10.80** | 14.20 | 5.54 | 7.65 | **4.06** | 5.82 | **11.43** |
| | **7.64** | 12.35 | 13.20 | 16.71 | **7.76** | 6.85 | **6.07** | 14.17 | |
| | 12.36 | 29.51 | **3.32** | 14.48 | 13.12 | 15.79 | **14.12** | 11.48 | |

It can be known from Table I that the average CIEDE2000 value of the color patches recovered by the algorithm in this paper is the smallest, indicating that the algorithm has better color recovery ability than other algorithms.

## 2. Definition evaluation

Through the gradient map experiment and the local detail comparison experiment, the effectiveness of the algorithm in the definition enhancement is verified from the overall and local details of the underwater image. The gradient image experiment is shown in Fig.9, and the local detail comparison experiment is shown in Fig.10.

It can be seen in Fig.9 that the gradient map of the enhanced image restores more overall details in terms of contrast and visibility.

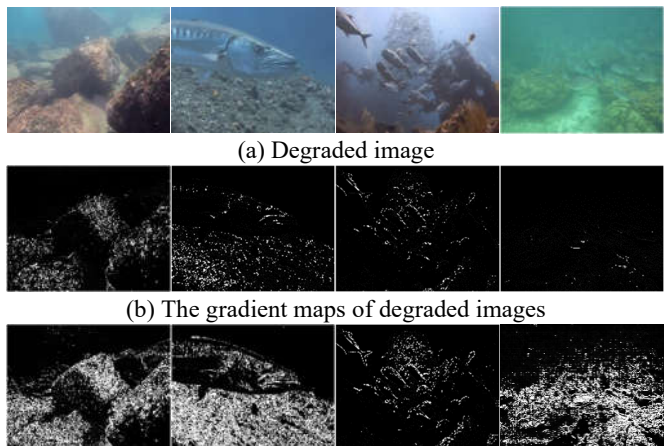

(a) Degraded image

(b) The gradient maps of degraded images

(c) The corresponding gradient maps of enhanced results

Fig. 9. Gradient map experiment.

The algorithm in this paper is compared with Underwater Dark Channel Prior (UDCP) [9] algorithm, fusion [2] algorithm and Image Color Correction Based on Double Transmission Underwater Imaging Model [10] algorithm, which can effectively enhance the definition. The red rectangular area is cut out from the test image for evaluation, as shown in Fig.10 (a). Fig.10 (b) shows the red rectangular area corresponding to each algorithm processing image.

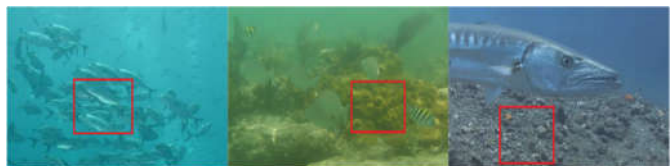

(a) Degraded image

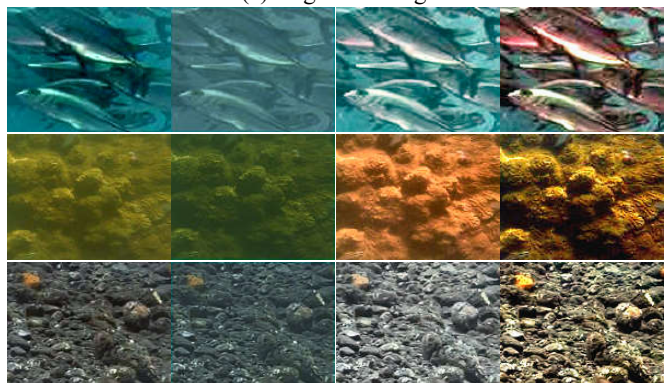

(b)　　(c)　　(d)　　(e)

Red rectangular area: (b) UDCP (c) Fusion (d) Ref. [10] (e) Our

Fig. 10. Local detail contrast experiment.

It can be seen in Fig.10 (e) that the contrast of our method image is enhanced. As shown in red rectangular area, the distinction between fish and water body is significantly improved, and the detail texture of water grass and rock is clearer.

## 3. Real underwater image evaluation

The algorithm and the comparison algorithm are applied to several representative real underwater images, these degraded underwater images are affected by color distortion, high turbidity and low contrast to varying degrees. The experimental recovery results of each algorithm are shown in Fig.11.

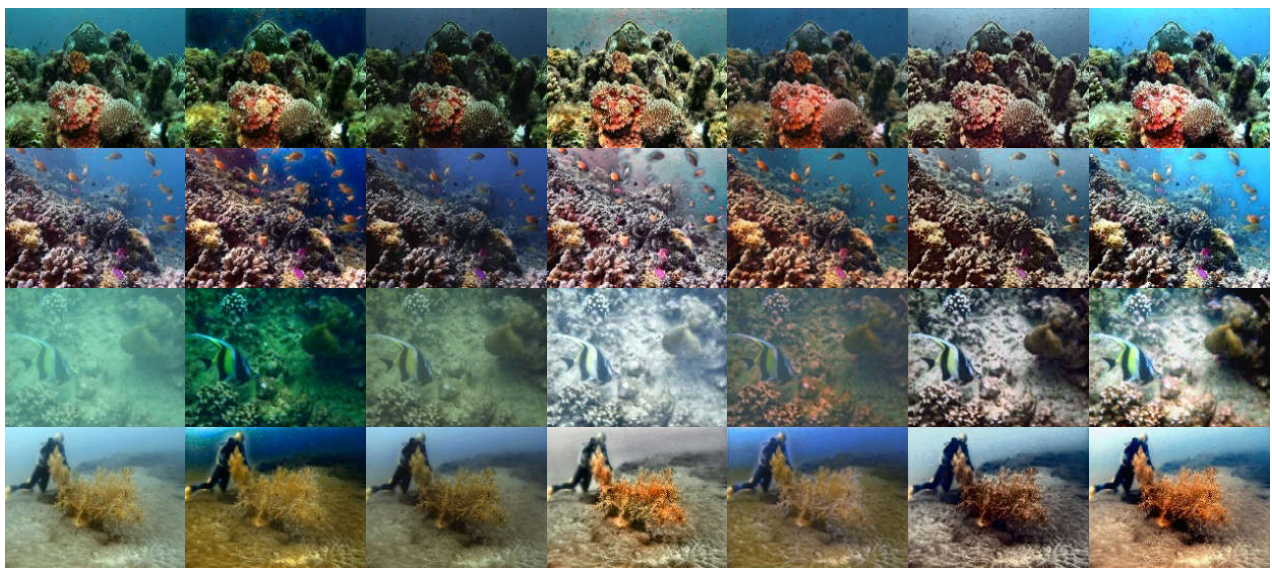

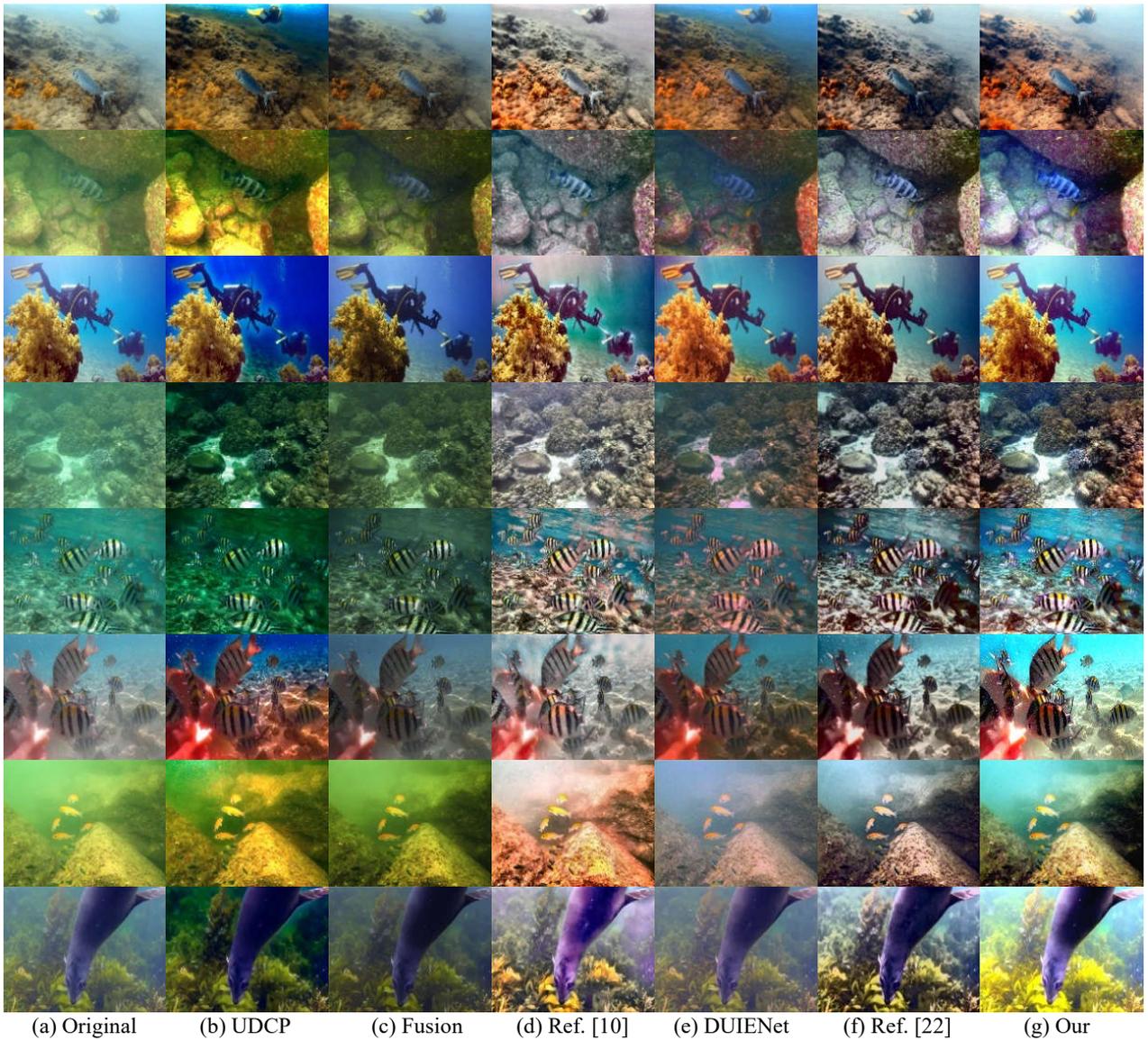

(a) Original    (b) UDCP    (c) Fusion    (d) Ref. [10]    (e) DUIENet    (f) Ref. [22]    (g) Our
Fig. 11. Real underwater image contrast experiment.

It can be seen that due to the lack of effective color correction processing in the UDCP algorithm, the resulting image has a problem of darkening the color. The details of the image processed by the fusion algorithm are relatively clear. Although the color is corrected, the color is still not vivid. Image Color Correction Based on Double Transmission Underwater Imaging Model [10] algorithm has a certain effect in enhancing image contrast, and the color is bright after correction, but the 11th image appears color distortion, and the algorithm robustness is slightly poor. Part of the image of the DUIENet algorithm is biased toward red, and red artifacts appear in these images. The algorithm has poor resilience, and a few images still have a fogging phenomenon. A few images of DehazeNet and HWD [22] algorithm are not bright in color, but the image clarity is effectively improved. Our algorithm performs well in all real underwater scenes. After processing, the image color is more natural, the color restoration conforms to the human visual effect, and can effectively restore the image visibility and detail information. Visibility in the far-sight area of the image is improved and the texture details in the near-field area are clearer. Compared with other algorithms, the algorithm in this paper effectively improves the visual quality of degraded images.

*B. Objective Evaluation*

Through subjective evaluation, we found that the algorithm has a good enhancement effect on underwater images with different color deviations and turbidity. Next, we further evaluate the image quality of the algorithm through four evaluation indexes.

1. The index of Underwater Color Image Quality Evaluation [24] (UCIQE) is used for quality evaluation, UCIQE uses the weighted combination of hue, saturation, and clarity of CIELAB space to evaluate the quality of the underwater color image. The value range of the UCIQE index is [0,1], which is directly proportional to the quality of the underwater color image. The higher evaluation value is, the better balance performance among hue, saturation, and clarity are.

2. The Perception-based Image Quality Evaluator [25] (PIQE) index calculates the No-Reference quality score of an image through the block by block distortion estimation, the value range of the PIQE index is [0,100]. The evaluation value is inversely proportional to the image quality. Low evaluation value indicates that the image has high quality, and a high evaluation value indicates that the image quality is low.

3. The Information entropy [5] can describe the information richness of the measured image, the clear image should have a larger information entropy value, which can be obtained by averaging the information entropy values of different channels.

4. The edge part of the image concentrates most of the information on the image. The Canny operator [26] is used to detect the edge of the image. The more complete edge details of the object in the image, the better the image effect, and more suitable for later analysis.

Table II, Table III and Table IV show the UCIQE evaluation index value, PIQE evaluation index value and Information entropy evaluation value of Fig.11, the bold font is the optimal value of the corresponding algorithm. Fig.12 is a comparison of Canny operator edge detection of three high turbidity underwater images. Because of the low definition of the image, edge detail detection is more challenging.

TABLE II. Evaluation of UCIQE index

| Methods | UDCP | Fusion | Ref. [10] | DUIENet | Ref. [22] | Our |
|---|---|---|---|---|---|---|
| Group-1 | 0.6627 | 0.5907 | 0.6559 | 0.6319 | 0.6164 | **0.6987** |
| Group-2 | 0.6670 | 0.5996 | 0.6135 | 0.6616 | 0.6262 | **0.6911** |
| Group-3 | 0.6000 | 0.4329 | 0.4577 | 0.4590 | 0.5698 | **0.6104** |
| Group -4 | 0.6613 | 0.5908 | 0.5962 | 0.5910 | 0.6257 | **0.6840** |
| Group -5 | 0.6676 | 0.5891 | 0.6103 | 0.6327 | 0.6103 | **0.6735** |
| Group -6 | 0.6429 | 0.5561 | 0.5254 | 0.5559 | 0.5713 | **0.6471** |
| Group-7 | 0.6735 | 0.6432 | 0.6662 | 0.6835 | 0.6541 | **0.7101** |
| Group-8 | 0.5710 | 0.4958 | 0.5252 | 0.5264 | 0.5771 | **0.6435** |
| Group-9 | 0.5494 | 0.5202 | 0.6369 | 0.5559 | 0.6138 | **0.6747** |
| Group-10 | 0.6972 | 0.5362 | 0.5655 | 0.5915 | 0.6389 | **0.6979** |
| Group-11 | 0.5906 | 0.5385 | 0.5899 | 0.4840 | 0.6130 | **0.6739** |
| Group-12 | 0.5755 | 0.5025 | 0.5959 | 0.5060 | 0.5832 | **0.6760** |
| Average | 0.6299 | 0.5496 | 0.5866 | 0.5733 | 0.6083 | **0.6734** |

TABLE III. Evaluation of PIQE index

| Methods | UDCP | Fusion | Ref. [10] | DUIENet | Ref. [22] | Our |
|---|---|---|---|---|---|---|
| Group-1 | 43.3795 | 41.4213 | 42.8095 | 62.0179 | 59.6470 | **38.5515** |
| Group -2 | 27.5039 | 29.5742 | 25.3489 | 62.3993 | 45.8692 | **24.8218** |
| Group -3 | 24.0762 | **16.3053** | 24.1734 | 67.1690 | 47.1704 | 22.5023 |
| Group -4 | 28.4479 | 29.3501 | 29.4175 | 66.5695 | 38.9655 | **28.2013** |
| Group -5 | 37.3341 | 39.5581 | 37.6146 | 77.5150 | 55.7698 | **28.5437** |
| Group -6 | 30.3624 | 29.5351 | 24.4300 | 65.0793 | 39.8712 | **14.2591** |
| Group-7 | 25.7824 | 27.9823 | **19.8579** | 66.4386 | 41.1657 | 24.8041 |
| Group-8 | 37.7465 | 34.7002 | 32.8571 | 65.4282 | 61.2821 | **23.6866** |
| Group-9 | **22.3647** | 24.2233 | 23.8514 | 68.8055 | 50.2152 | 24.9101 |
| Group-10 | 33.3224 | 34.6680 | 30.3886 | 72.9471 | 53.0191 | **23.1321** |
| Group-11 | 48.8929 | 42.2950 | 40.2336 | 42.1398 | 59.7618 | **27.7526** |
| Group-12 | 41.9371 | 34.5823 | 27.5674 | 33.4203 | 61.1796 | **26.1095** |
| Average | 33.4292 | 32.0163 | 29.8792 | 62.4941 | 51.1597 | **25.6062** |

TABLE IV. Evaluation of Information entropy index

| Methods | UDCP | Fusion | Ref. [10] | DUIENet | Ref. [22] | Our |
|---|---|---|---|---|---|---|
| Group-1 | 7.1810 | 6.8762 | 7.6236 | 7.3081 | 7.7241 | **7.7544** |
| Group -2 | 7.5468 | 7.2536 | **7.8687** | 7.4954 | 7.7768 | 7.8303 |

| Group -3 | 6.5386 | 6.8399 | 7.4328 | 6.4727 | 7.8082 | **7.8774** |
| Group -4 | 7.3313 | 7.3806 | 7.6296 | 7.2790 | 7.8174 | **7.8708** |
| Group -5 | 7.3643 | 7.5256 | 7.7152 | 7.4229 | 7.8194 | **7.9380** |
| Group -6 | 7.3843 | 6.9113 | 7.3187 | 6.7476 | 7.7299 | **7.8441** |
| Group-7 | 7.1434 | 7.7207 | 7.8218 | 7.6869 | 7.7687 | **7.9647** |
| Group-8 | 6.8446 | 6.6816 | 7.4887 | 6.6993 | 7.6461 | **7.6678** |
| Group-9 | 6.2226 | 6.8242 | 7.7242 | 6.9327 | 7.6908 | **7.8968** |
| Group-10 | 7.3171 | 6.9757 | 7.5619 | 7.0576 | **7.8172** | 7.5391 |
| Group-11 | 7.7215 | 7.5171 | 7.7751 | 7.0788 | 7.8283 | **7.8948** |
| Group-12 | 6.7148 | 6.6920 | 7.6920 | 6.8061 | **7.8432** | 7.7504 |
| Average | 7.1093 | 7.0999 | 7.6377 | 7.0823 | 7.7725 | **7.8191** |

It can be seen from the data in each table that the comparison results of multiple evaluation indexes verify the superiority of the algorithm in this paper applied to the real underwater image, which shows that the algorithm can better balance the hue, saturation, and clarity of underwater image, present more effective information and remove the atomization phenomenon of the underwater image well, with higher visual effect.

It can be seen from the Fig.12 that the edge details of the algorithm in this paper and UDCP algorithm are well maintained, while some details of other algorithms are slightly missing, but as shown in the red marking area, UDCP algorithm tends to maintain the background details, the algorithm in this paper effectively enhances the edge details of the main objects in the figure, such as the pipes and fish in the figure, while the important information in the image is mostly concentrated in the main objects, so algorithm performs better in edge detection experiments.

Finally, the feature point matching test is performed by speeded up robust features [27], comparing with the Underwater Dark Channel Prior (UDCP) [9] algorithm, fusion [2] algorithm, Image Color Correction Based on Double Transmission Underwater Imaging Model [10] algorithm, Deep Underwater Image Enhancement Network (DUIENet) [21] algorithm, Underwater Image De-Scattering and Enhancing Using DehazeNet and HWD (DehazeNet and HWD) [22] algorithm. The degraded image for testing is shown in Fig.13. Table V shows the number of feature point matching of the image before and after processing. The bold font is the optimal value of the corresponding algorithm. The first group image in Fig.13 is taken as an example. The example of feature point matching is shown in Fig.14.

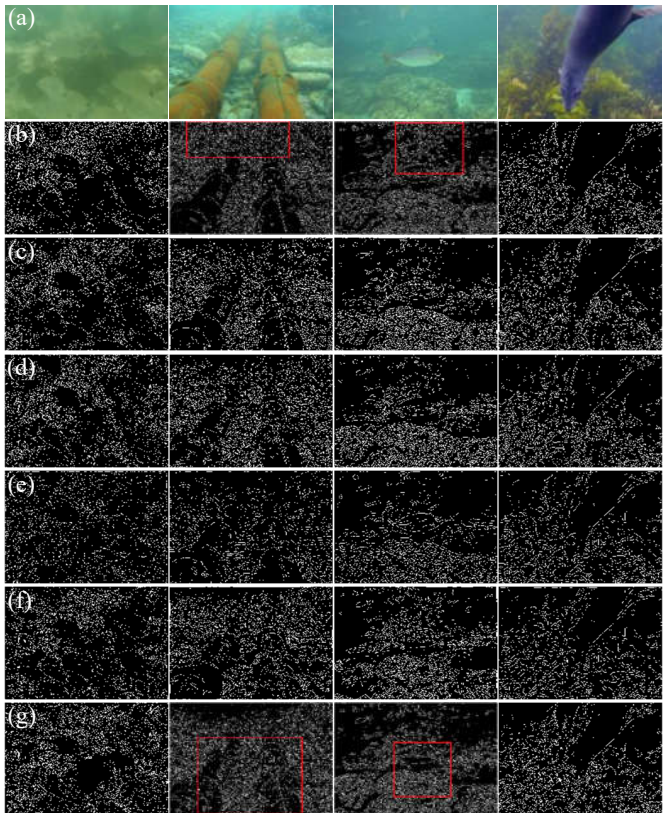

(a) Original  (b) UDCP  (c) Fusion  (d) Ref. [10]  (e) DUIENet
(f) Ref. [22]  (g) Our
Fig. 12. Comparison of canny operator edge detection.

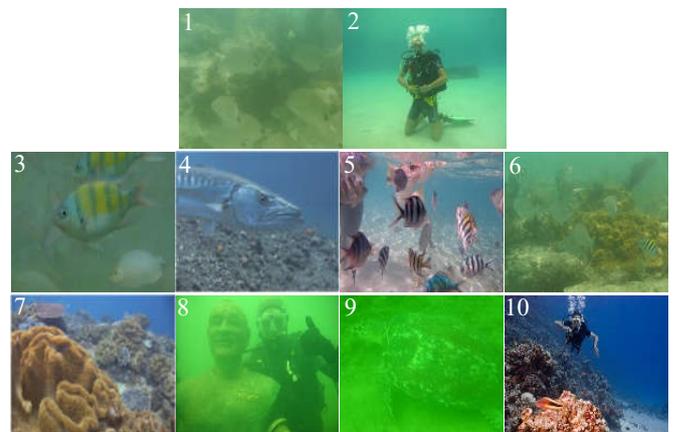

Fig. 13. Degraded image.

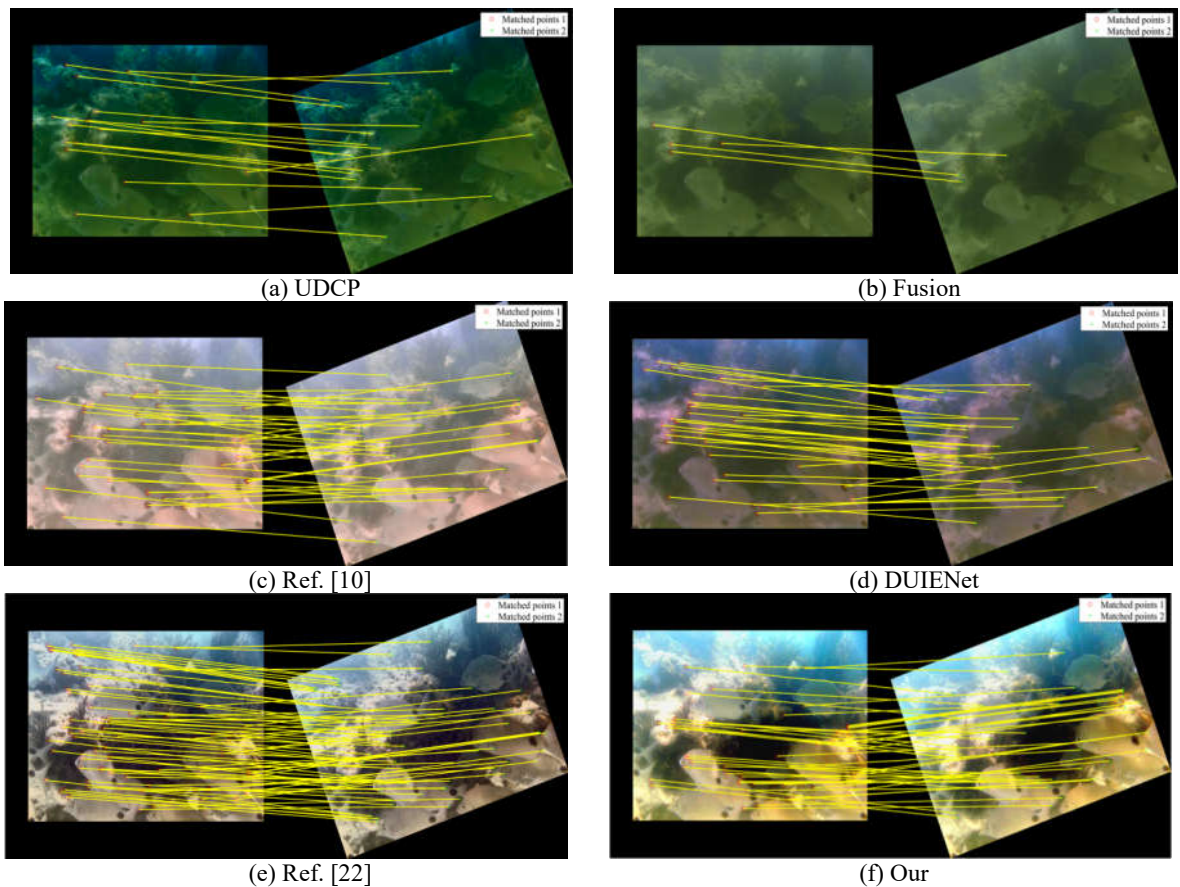
Fig. 14. Example of feature point matching.

TABLE V. Number of characteristic matching point

| Methods | 1 | 2 | 3 | 4 | 5 | 6 | 7 | 8 | 9 | 10 | Avg |
|---|---|---|---|---|---|---|---|---|---|---|---|
| UDCP | 19 | 23 | 15 | 13 | 32 | 30 | 47 | 20 | 8 | 13 | 22 |
| Fusion | 4 | 18 | 7 | 15 | 29 | 10 | 23 | 12 | 30 | 6 | 14.3 |
| Ref. [10] | 41 | 34 | 45 | 29 | 52 | 61 | 50 | 36 | 47 | 21 | 41.6 |
| DUIENet | 35 | 35 | 14 | **57** | **78** | 60 | 49 | 17 | 23 | **77** | 44.5 |
| Ref. [22] | **79** | 27 | **112** | 41 | 71 | **75** | **82** | 39 | **86** | 54 | **66.6** |
| Our | 50 | **36** | 51 | 28 | 62 | 55 | 64 | **39** | 43 | 22 | 45 |

It can be seen from the data in Table V that the number of feature matching points of the algorithm image in this paper is less than the number of feature matching points of the DehazeNet and HWD [22] algorithm. It can be seen from Fig.14 (e), (f) that although the algorithm in this paper effectively solves the problem of color distortion, the brightness of some areas of the image increases slightly, resulting in the inability to detect matching feature points. How to effectively control the brightness increase will be the focus of future research. Compared with other comparison algorithms, this algorithm has more feature matching points, which shows that this algorithm can effectively restore the clarity of the image, and has a good effect in the following feature matching process.

V. CONCLUSION

To solve the problems of color distortion, blurring and excessive noise of the underwater image, an Underwater Image Enhancement Based on Structure-Texture Reconstruction is proposed. The algorithm aims at the problem that the degraded image contains too much texture information and the amount of noise that can't be ignored, which makes the estimation of transmittance image unreliable. The degraded image is decomposed into the structural layer and texture layer, and the transmittance image is well estimated from the structural layer, which greatly removes the influence of interference factors. RDCP algorithm is used to optimize the transmittance of the backscatter component, and the background light value is calculated accurately under the worse imaging conditions. The solution is completed by the underwater optical imaging model, which effectively solves the problems of color deviation and blurring caused by scattering in the structure image under the natural light conditions, through multi-scale detail lifting algorithm and binary mask to enhance the effective details of texture image and remove the interference artifacts, the enhanced texture layer is added back to the original structure layer to generate the final result image. Experimental results show that compared with the mainstream UDCP algorithm and the novel DehazeNet and HWD, DUIENet, fusion underwater image clarity algorithms, our method better balances the hue, saturation and sharpness of the image, the color information of the resulting image is recovered naturally and the texture details are clear. But the brightness of a few images is too high, how to optimize will be the direction of future research.